# WHENCE TYCHO'S CASE AGAINST COPERNICUS? ON GENESIS, AUGUSTINE, AND THE STARS

Christopher M. Graney
Vatican Observatory
00120 Stato Città del Vaticano
c.graney@vaticanobservatory.org

**ABSTRACT:**

**This paper argues that Tycho Brahe's "principal argument against Copernicus" (as the astronomer Christiaan Huygens called it) likely derived from a much older argument regarding the sizes of the "two great lights" described in the first chapter of the book of Genesis. Brahe's argument, that in the Copernican system stars would have to be absurdly large, played an important role in opposition to the Copernican system in the seventeenth century. Brahe presented the argument in an exchange of letters with Christoph Rothmann in 1588-89. Within that exchange Rothmann and Brahe touched both on the question of the two great lights of Genesis and on theologians such as Augustine of Hippo who treated that question. The fundamentals of Brahe's important line of argument against Copernicus thus well pre-dated Copernicus and Brahe.**

**INTRODUCTION:**

At the end of the seventeenth century, the astronomer Christiaan Huygens (1629-95) outlined what he called Tycho Brahe's "principal argument" against the heliocentric world system of Copernicus:

> Before the invention of telescopes, it seemed to contradict Copernicus's opinion, to make the sun one of the fixed stars. For the stars of the first magnitude being esteemed to be about three minutes diameter; and Copernicus (observing that though the earth changed its place, they always kept the same distance from us) having ventured to say that the *magnus orbis* [Earth's orbit] was but a point in respect of the sphere in which they were placed, it was a plain consequence that every one of them that appeared any thing bright, must be larger than the path or orbit of the earth: which is very absurd. This is the principal argument that Tycho Brahe set up against Copernicus. But when the telescopes took away those rays of the stars which appear when we look upon them with our naked eye, (which they do best when the eye-glass is blacked with smoke) they seemed just like little shining points, and then that difficulty vanished, and the stars may be so many suns.[1]

From where did Brahe (1546-1601) get this star size argument? The argument played a significant role in opposition to the Copernican system, and carried weight at least into the later seventeenth century. This paper proposes that Brahe obtained the idea for this argument from a thousand-year-old discussion regarding the sizes of celestial bodies and the first chapter of Genesis.

Brahe might have come upon this discussion in the works of Augustine of Hippo (354-430), Thomas Aquinas (1225-74), or John Calvin (1509-64). The most likely source among these three is Augustine of Hippo, whose writings Brahe mentions in the same letter in which he presents the star size argument. Even if Brahe did not derive his idea from one of these writers, it is clear from the Genesis discussion that the basic outlines of Tycho Brahe's principal argument well predated Brahe.

**GENESIS AND THE STARS, WITH PTOLEMY, AUGUSTINE, AQUINAS AND CALVIN:**
Genesis 1:14-16 describes the creation of the sun, moon, and stars, and in doing so alludes to their sizes:

> And God said, Let there be lights in the firmament of heaven to divide the day from the night; and let them be for signs, and for seasons, and for days and years: And let them be for lights in the firmament of heaven to give light upon the earth: and it was so. And God made two great lights; the greater light to rule the day, and the lesser light to rule the night: *he made* the stars also.[2]

If the sky is a dome with these lights on its surface, then the "greatness" of these lights is a matter of simple sight. The sun, moon, and stars are all the same distance from Earth; therefore the relative physical sizes of these bodies are just as they appear to the eye. The sun and moon appear larger than the stars, and so they are larger in terms of actual physical bulk as well.

**Ptolemy and Genesis:**
Careful study of the sky, however, reveals it not to be a dome. Ptolemy (~150) in his *Almagest* discussed how observations of the stars showed the Earth to be merely a point in comparison to the distance to the stars:

---

[1] Christiaan Huygens, *The Celestial Worlds Discover'd* (London, 1722), 145. *The Celestial Worlds* is a translation of Huygens' 1698 *Kosmotheoros*.
[2] King James Version, appropriate for the time period.

> Now, that the earth has sensibly the ratio of a point to its distance from the sphere of the so-called fixed stars gets great support from the fact that in all parts of the earth the sizes and angular distances of the stars at the same times appear everywhere equal and alike, for the observations of the same stars in the different latitudes are not found to differ in the least.[3]

By contrast, observations of the moon from different places on the Earth's globe do differ, showing that Earth is not merely a point in comparison to the distance to the moon (Figure 1).

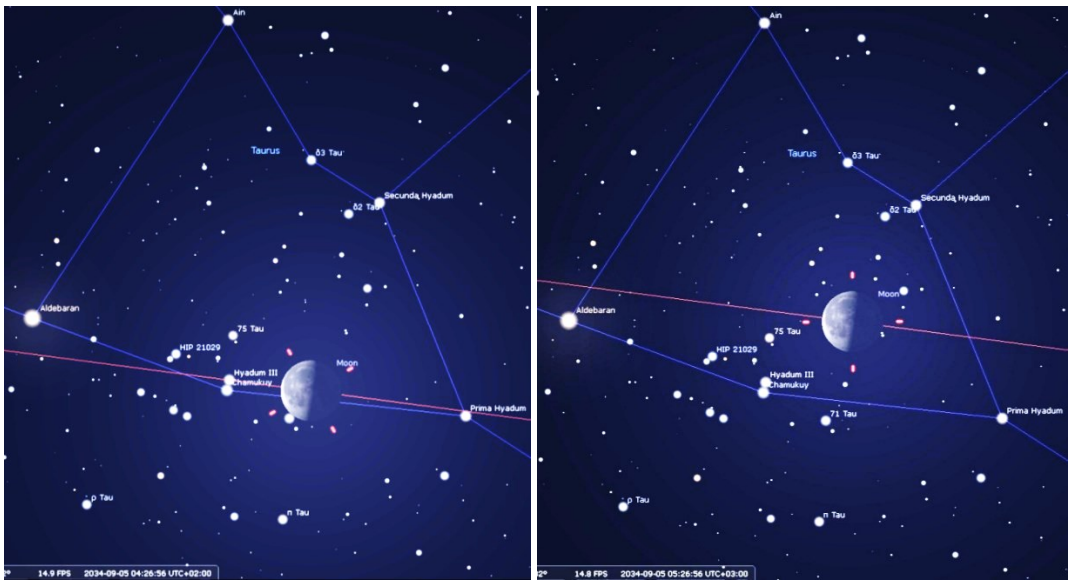

**Figure 1: The moon passing through the Hyades in Taurus on 5 September 2034, as seen from Gdansk (left) and Khartoum. The red line is the moon's path through the stars. Observations of the same stars from different latitudes do not differ, indicating that the stars are so distant that the size of the Earth is as nothing by comparison, but this is not true in the case of the moon. Images made with the planetarium software *Stellarium*.**

Ptolemy's science was persuasive. Anyone who travelled and had good eyesight could confirm what Ptolemy said. Thus despite the contradiction between the domed universe described by Genesis and the calculations of Ptolemy, the Christian writer Severinus Boethius would cite Ptolemy by name in his *On the Consolation of Philosophy* (523) and write,

---

[3] Ptolemy, "The Almagest I, 6" in *Great Books of The Western World (16): Ptolemy, Copernicus, Kepler* (Chicago: W. Benton, 1952), 10.

> You have learned from astronomy, that this globe of earth is but as a point, in respect to the vast extent of the heavens; that is, the immensity of the celestial sphere is such that ours, when compared with it, is as nothing, and vanishes.[4]

Ptolemy determined the stars to have small but measureable apparent sizes (apparent diameters of a couple of minutes of arc—roughly one fifteenth the apparent diameter of the moon—in line with the statement of Huygens above). The vast distance to the stars meant that they had to be very large to appear even that small in the night sky. Ptolemy determined the actual diameter of the most prominent stars to be more than four times that of Earth. He calculated that the sun was actually five times Earth's diameter, and thus twenty-five times its bulk, while the moon was actually less than one third of Earth's diameter.[5]

A prominent star was therefore far greater than the moon (Figure 2). Indeed, every visible star in the night sky would greatly exceed the moon in terms of bulk. Ptolemy's calculations showed that the moon hardly qualified as one of the great heavenly lights. Anyone with good eyesight who cared to look could at least approximately confirm Ptolemy's measurements of the apparent sizes of the stars compared to the apparent size of the moon. The stars might *appear* small, but the moon *was* small. The moon was arguably not a "great light". This conflict with Genesis is what brought the star size question to the attention of Augustine, Aquinas, and Calvin.

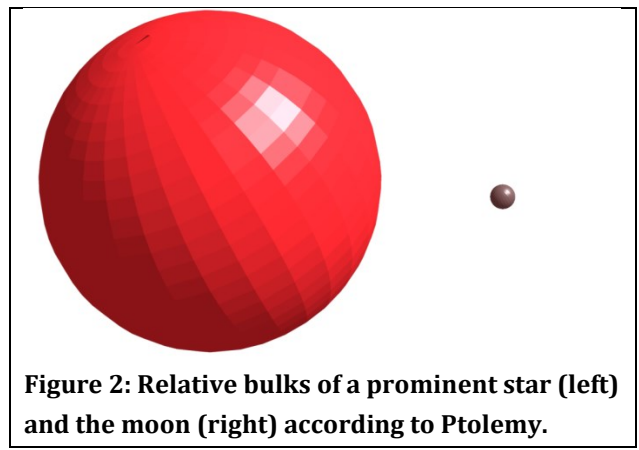

Figure 2: Relative bulks of a prominent star (left) and the moon (right) according to Ptolemy.

**Augustine and Genesis:**

Augustine discussed the issue of star sizes and distances and Genesis in his *On the Literal Interpretation of Genesis*:

> A question also commonly asked is whether these conspicuous lamps in the sky, that is, sun and moon and stars, are all equally brilliant, but because of their different distances from the earth appear to our eyes for that reason to vary in brightness. And about the moon those who take this line do not hesitate to say that its light is in itself less than that of the sun, by which they also maintain it is illuminated. Many of the stars, however, so they boldly assert, are equal to the sun, or even greater, but they seem small because they have been set further away.[6]

---

[4] Philip Ridpath (trans.), *Boethius's Consolation of Philosophy* (London: C. Dilly, 1785), 67.
[5] Albert Van Helden, *Measuring the Universe: Cosmic Dimensions from Aristarchus to Halley* (Chicago: University of Chicago Press, 1985), 27

After further elaboration on what might be said about the celestial lights, Augustine concluded:

But let them say what they like about the heavens, those who are strangers to the Father who is in heaven. As for us, though, it is not our business to inquire more precisely into the size and the spacing of the constellations, spending time on such things that is needed for more important and worthy matters. We even think it more worthwhile to attend to those lights that are greater than the rest, which scripture draws to our attention as follows: *And God made two great lights*, which for all that are not equal. It goes on to say, you see, after putting them ahead of the rest, that they differ from each other. It says, I mean: *the greater light for the starting of the day, and the lesser light for the starting of the night* (Gn 1:16). Let them at least grant this to our eyes, after all, that it is obvious that they shine more brightly than the rest upon the earth, and that it is only the light of the sun that makes the day bright, and that even with so many stars appearing, the night is never as light when there is no moon, as when it is being illuminated by its presence.[7]

**Aquinas and Genesis:**

Aquinas addressed the "great lights" question in his *Summa Theologica*, Question LXX ("Of the Work of Adornment, as regards the Fourth Day—In Three Articles"). He considers various objections to the Genesis account regarding the lights, including,

*Obj. 5.* Further, as astronomers say, there are many stars larger than the moon. Therefore the sun and the moon alone are not correctly described as the *two great lights*.[8]

His answer to this:

*Reply Obj. 5.* As Chrysostom says, the two lights are called great, not so much with regard to their dimensions as to their influence and power. For though the stars be of greater bulk than the moon, yet the influence of the moon is more perceptible to the senses in this lower world. Moreover, as far as the senses are concerned, its apparent size is greater.[9]

---

[6] St. Augustine (Bishop of Hippo), *The Literal Meaning of Genesis*, in *The Works of Saint Augustine, a Translation for the 21st Century, Part I, Volume 13: On Genesis: A Refutation of the Manichees; Unfinished Literal Commentary on Genesis; The Literal Meaning of Genesis—introductions, translation and notes by Edmund Hill, O.P., editor John E. Rotele, O.S.A.* (Hyde Park, NY: New City Press, 2002), 211 (Book II, 16.33).
[7] *The Literal Meaning of Genesis*, 212
[8] *"The Summa Theologica" of St. Thomas Aquinas, Part I. QQ. I-LXXIV. Literally Translated by Fathers of the English Dominican Province, Second and Revised Edition* (London: Burns, Oates & Washbourne, Ltd., 1922), 238-239. *"The Summa Theologica"*, 239.

**Calvin and Genesis:**

Calvin made the same general point as Augustine and Aquinas regarding the question of star sizes and Genesis, but at greater length:

> Moses makes two great luminaries; but astronomers prove, by conclusive reasons that the star of Saturn, which on account of its great distance, appears the least of all, is greater than the moon. Here lies the difference; Moses wrote in a popular style things which without instruction, all ordinary persons, endued with common sense, are able to understand; but astronomers investigate with great labor whatever the sagacity of the human mind can comprehend.... [B]ecause [Moses] was ordained a teacher as well of the unlearned and rude as of the learned, he could not otherwise fulfill his office than by descending to this grosser method of instruction. Had he spoken of things generally unknown, the uneducated might have pleaded in excuse that such subjects were beyond their capacity. Lastly since the Spirit of God here opens a common school for all, it is not surprising that he should chiefly choose those subjects which would be intelligible to all. If the astronomer inquires respecting the actual dimensions of the stars, he will find the moon to be less than Saturn; but this is something abstruse, for to the sight it appears differently. Moses, therefore, rather adapts his discourse to common usage.... [Moses] does not call us up into heaven, he only proposes things which lie open before our eyes.[10]

Calvin here chose for his example the "star" Saturn rather than a "fixed" star, but since Ptolemy had the fixed stars located just beyond Saturn, and since to the eye Saturn looks like many a star, the point is the same—one of the apparently small lights in the sky must actually exceed the moon in terms of true bulk.

Calvin also discussed this in his commentary on Psalm 136, which in verses 7-9 also speaks of the great lights:

> O give thanks to the LORD...
> To him that made great lights:
> for his mercy *endureth* forever.

---

[9] *"The Summa Theologica"*, 242.
[10] John Calvin, "Chapter 1" in *Commentaries on the First Book of Moses, called Genesis, by John Calvin, Vol. 1, John King, trans.* (Edinburgh: Calvin Translation Society, 1847) par. 16, 86-87.

> The sun to rule by day:
> for his mercy *endureth* forever.
> The moon and stars to rule by night:
> for his mercy *endureth* forever.[11]

Regarding these verses, Calvin wrote:

> Moses [in Genesis 1:16] calls the sun and moon the two great lights, and there is little doubt that the Psalmist here borrows the same phraseology. What is immediately added about the stars, is, as it were, accessory to the others. It is true, that the other planets are larger than the moon, but it is stated as second in order on account of its visible effects. The Holy Spirit had no intention to teach astronomy; and, in proposing instruction meant to be common to the simplest and most uneducated persons, he made use by Moses and the other Prophets of popular language, that none might shelter himself under the pretext of obscurity, as we will see men sometimes very readily pretend an incapacity to understand, when anything deep or recondite is submitted to their notice. Accordingly, as Saturn though bigger than the moon is not so to the eye owing to its greater distance, the Holy Spirit would rather speak childishly than unintelligibly to the humble and unlearned.[12]

All three writers accept that the stars are larger than the moon in terms of actual size. All emphasize that the Bible is speaking to appearances when it calls the moon a "great light". Augustine notes that "to our eyes" the sun and moon are the two great lights; they illuminate the Earth whereas the stars do not. Aquinas says that Genesis is speaking of the "apparent size" of the moon. Calvin says that the Bible is describing what is "visible", what "lies open before our eyes"; the Bible is not written to "teach astronomy" or to "call us up into heaven" and show us the actual sizes of the moon and stars. Other writers also treated the question of star sizes and the Bible.[13]

**BRAHE'S PRINCIPLE ARGUMENT**

Under the Copernican world system, the question of star sizes and the Bible readily transforms into Brahe's principal argument. If Earth circles the sun, then Earth's *orbit* replaces Earth's *globe* in Ptolemy's argument about the distance to the stars. Observations of the same stars from different locations on Earth's *orbit* (that is, at different times of the year) are not found to differ in the least, so Earth's *orbit* is like a point compared to the distance to the stars. A prominent star then becomes, not four times the diameter of *Earth*, and thus a little smaller than the sun, like Ptolemy had calculated, but four times the diameter of *Earth's orbit*. That prominent star now utterly dwarfs the sun. For that matter, so does every star seen in the night sky, even the least prominent. Such gigantic stars are, in the words of Huygens (and of Brahe, as we shall see), "absurd". End of argument.

---

[11] KJV.

[12] John Calvin, "Psalm CXXXVI" in *Commentary on the Book of Psalms, by John Calvin, Vol. 5*, James Anderson, trans. (Edinburgh: Calvin Translation Society, 1849) par. 7, 184-185.

[13] Aquinas notes John Chrysostom from the fourth century, although the reference is unclear. Another is Andrew Willet, *Hexapla in Genesin, that is, a Sixfold Commentary upon Genesis, etc.* (London, 1605) who discusses the smallness of the moon versus the stars on page 10.

Star sizes and the Bible may readily transform into Brahe's argument, but is that how Brahe obtained the argument? There is nothing to indicate unambiguously that he did. Nowhere does he say, for example, that his principal argument can be derived from what Aquinas says in *Summa Theologica*, Question LXX. Indeed, Brahe never mentions Aquinas at all in his work.[14]

Brahe does mention Calvin, in a 1590 letter to Caspar Peucer, and there in regards to the first chapter of Genesis. However, his discussion pertains to Genesis 1:6-7 and the waters "above the firmament".[15] This provides reason to think that Brahe was familiar with Calvin's thoughts on Genesis, but that is all.

Brahe touches on Augustine more, and he discusses the two great lights of Genesis. These come in an exchange of letters in 1588-89 between Brahe and Christoph Rothmann (d. ~1600), part of their discussion on the Copernican system in which Brahe says he persuaded Rothmann to abandon Copernicanism.[16] Rothmann had brought up the question of Genesis and the sizes of stars in a letter of 19 September 1588, in which he was advocating for a free interpretation of the Bible on matters of astronomy:

> The scriptures are not written only for you and me, but for all humanity altogether; indeed, they speak to the comprehension of all, as indeed all theologians acknowledge in explications of Genesis Chapter 1. Otherwise the moon might be greater than the rest of the stars, contrary to geometrical demonstrations. But from those, that is clear.[17]

---

[14] J. L. E. Dreyer, *Tycho Brahe Dani Opera Omnia [TBOO] XV: Index Hominum et Rerum* (Copenhagen, 1929), 11.
[15] *TBOO VII: Epistolae Astronimicae Tomus II* (1926), 233.
[16] Peter Barker, "How Rothmann Changed His Mind", *Centaurus* 46:1 (2004), 41-57. Bernard Goldstein and Peter Barker, "The Role of Rothmann in the Dissolution of the Celestial Spheres", *The British Journal for the History of Science* 28:4 (1995), 385-403. Robert Westman, *The Copernican Question: Prognostication, Skepticism, and Celestial Order* (Oakland, California: University of California Press, 2011), 290-99.
[17] *TBOO VI*, 159: "Hae enim non mihi & tibi solummodo, verum omnibus omnino hominibus scriptae sunt, ad quorum captum etiam loquuntur, vt etiam omnes theologi in explicatione capitis 1. Genes. fatentur. Alias Luna contra Geometricas demonstrationes esset maior reliquis Stellis. Sed & illud ex hoc manifestum est." See also Scott Mandelbrote and Jitse van der Meer, *Nature and Scripture in the Abrahamic Religions: Up to 1700 (Vol. I)* (Leiden: Brill, 2008), 571.

Rothmann had also brought up Augustine, at the start of a letter of 22 August 1589 which he opened by defending himself from what he saw as Brahe's accusation that he was being too free with the words of scripture.[18] "If you will read Augustine," Rothmann told Brahe, "you will discover much freer speaking concerning the sacred writings."[19]

One instance in which Brahe had suggested that Rothmann was being too free with scripture was in a 21 February 1589 letter, where Brahe had discussed Genesis 1 and the matter of the two great lights.[20] Here Brahe had mentioned that "Moses" (Genesis) called the sun and moon "great" based on "magnitude of light", not on "size of bodies". Brahe had also mentioned the vast distances of the stars, and he had noted Moses's need to not delve into the details of astronomy, "for he is writing to a rude people"—thus agreeing with Rothmann.[21] But Brahe told Rothmann, "You disparage the Prophets too much, when you assert that they do not understand more than the common people concerning the nature of things."[22]

Brahe responded to Rothmann's August letter in a letter of 24 November, which he opens both by assuring Rothmann that he was never questioning his piety,[23] and also by again drawing attention to Moses and the magnitudes of the moon and stars.[24] Then he touches on Augustine. Augustine speaks freely enough, he says, but not in a way that bears on the Copernican question. Brahe goes on to say that if Rothmann has anything from Augustine or any other religious authority that is in favor of the Copernican assertion, he should cite the writings in question; Brahe also mentions that Augustine denied the possibility of antipodes.[25] Quite a bit later in this lengthy letter, Brahe again brings up Moses and the two great lights. He mentions Augustine in the same (lengthy) sentence. However, the mention is regarding theologians and the motion of the Earth, and not about those lights (and Brahe does not cite a specific writing).[26]

---

[18] Adam Mosley, *Bearing the Heavens: Tycho Brahe and the Astronomical Community of the Late Sixteenth Century* (Cambridge: Cambridge University Press, 2007), 97.

[19] *TBOO VI*: *Epistolae Astronomicae Tomus I* (1919), 181: "Si Augustinam legeris, multo liberius de sacris literis loquentem invenies." See also Ann Blair, "Tycho Brahe's Critique of Copernicus and the Copernican System", *Journal of the History of Ideas* Vol. 51, No. 3 (1990), 355-377, on 363, and Goldstein and Barker, 398.

[20] Blair, "Tycho Brahe's Critique", 363.

[21] J. L. E. Dreyer, *Tycho Brahe: A Picture of Scientific Life and Work in the Sixteenth Century* (Edinburgh, 1890), 177. *TBOO VI*, 177: "Hic saltem de magnitudine lucis, qua Sol & Luna prae caeteris Stellis Terras illuminant, loquitur Moses, non de quantitate corporum."; "licet ob nimiam distantiam lux illa Terras non percellat. Nam & Moses eo loco de hoc saltem lumine agebat quod Terras contingit"; "Sic Moses etsi in primo capite Geneseos de Mundi creatione agens, Astronomiae penetralia non reseret, vtpote rudi populo scribens."

[22] *TBOO VI*, 177: "nimium Prophetis derogas, dum ais ipsos non plus intellexisse de rerum natura, quam alios vulgares homines". See Blair, 363.

[23] Blair, 363. *TBOO VI*, 185.

[24] *TBOO VI*, 185.

[25] Blair, 363. *TBOO VI*, 186.

[26] *TBOO VI*, 195.

Two pages later Brahe lays out his Principle Argument. The fact that observations of the same stars from different locations on Earth's orbit are not found to differ, he says, puts a minimum distance on the stars—seven hundred times the distance to Saturn. Given that,

> Then stars of the third magnitude which are one minute in [apparent] diameter will necessarily be equal to the entire annual orb [Earth's orbit], that is, they would comprise in their diameter 2,284 semidiameters of the earth. They will be distant by about 7,850,000 of the same semidiameters. What will we say of the stars of first magnitude, of which some reach two, some almost three minutes of visible diameter? and what if, in addition, the eighth sphere were removed higher,[27] so that the annual motion of the earth vanished entirely [and was no longer perceptible] from there? Deduce these things geometrically if you like, and you will see how many absurdities (not to mention others) accompany this assumption [of the motion of the Earth] by inference.[28]

**CONCLUSION**

Nowhere does Brahe directly indicate that this, his "principal argument" (as Huygens called it), derives from what Calvin and Aquinas say regarding Genesis 1 and the "two great lights". However,

- Brahe was familiar with Calvin on Genesis 1.
- He was familiar with the question of astronomical observations yielding results contrary to the plain words of Genesis regarding those lights.
- He was familiar with Augustine.
- He had discussed both Augustine and (separately) the two great lights question in the same letter to Rothmann in which he put forth his principal argument—once in the same sentence.

This could all be a matter of coincidence. The case for coincidence, however, seems weaker than the case for the principal argument being derived from the work of Calvin and Augustine and the matter of the "two great lights".

Brahe's principal argument against the Copernican system was very influential. While Huygens described it as having force before the invention of telescopes, in fact it had force for decades after telescopes came on the scene, too. Telescopes in the first half of the seventeenth century had very small apertures. Owing to the wave nature of light

---

[27] That is, if observations were improved and yet did not yield a detection of differences between different locations on Earth's orbit, so that the minimum distance to the stars must become greater.

[28] *TBOO VI*, 197. Dreyer, *Tycho Brahe*, 176-77. Blair, 364 (translation from Blair).

(diffraction), such telescopes formed images of stars that had the appearance of globes whose diameters, while smaller than what was seen by the unaided eye, were nevertheless measurable (Figure 3). Telescopes also increased the sensitivity of observations, forcing, as Brahe had imagined, the stars to be still farther removed from Earth. The result was that the principal argument remained. Anti-Copernican telescopic astronomers such as Christoph Scheiner, Simon Marius, Giovanni Battista Riccioli, and others wielded the argument against Copernicus. Copernican astronomers such as Johannes Kepler and Philips Lansbergen simply accepted the argument, rejecting instead the idea that stars having sizes "larger than the path or orbit of the earth… is very absurd". Well into the eighteenth century, respected astronomers continued to argue that telescopic measurements showed stars to be of such sizes. In due course, discoveries such as that mentioned by Huygens—that the image of any star significantly shrinks when viewed through a filter such as a smoked glass, and that therefore star images such as that seen in Figure 3 are spuriously large and do not reflect the actual physical globe of the star (a discovery Huygens first mentioned in 1659)—dissolved Tycho Brahe's principal argument against the Copernican system.[29]

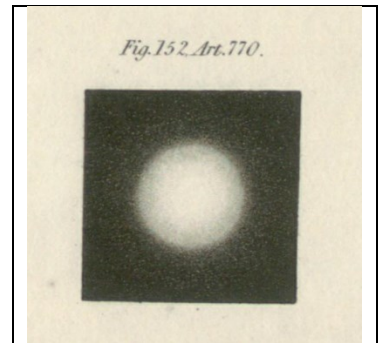

**Figure 3: A star as seen through a telescope of very small aperture, as illustrated in John Herschel's 1828 *Treatise on Physical Astronomy, Light and Sound* (Plate 9). The globe-like appearance is a spurious product of the wave nature of light; it does not reflect the physical globe of the star, and greatly inflates the apparent size of the star. Christiaan Huygens discovered that filtering the light of the star shrank this globe, revealing its spurious nature. Image credit: ETH-Bibliothek Zürich, Alte und Seltene Drucke.**

But before such discoveries, Brahe's argument regarding the sizes of stars was a powerful objection to the Copernican system, used by many anti-Copernican astronomers during the seventeenth century. It seems likely that Brahe derived this argument from discussions on Genesis 1 and the sizes of celestial bodies. These discussions dated back over a thousand years, to the work of Augustine of Hippo. They certainly contained the outlines of Brahe's important argument.

---

[29] This material is treated in Christopher M. Graney, *Setting Aside All Authority: Giovanni Battista Riccioli and the Science Against Copernicus in the Age of Galileo* (Notre Dame: University of Notre Dame Press, 2015), except for the ideas of Kepler, the persistence of the star size question into the eighteenth century, and the work of Scheiner. For these, see Graney, "The Starry Universe of Johannes Kepler", *Journal for the History of Astronomy* 50:2 (2019), 155-73; "The Starry Universe of Jacques Cassini: Century-old Echoes of Kepler", *Journal for the History of Astronomy* 52:2 (2021), 147-67; and *Mathematical Disquisitions: The Booklet of Theses Immortalized by Galileo* (Notre Dame: University of Notre Dame Press, 2017), respectively. For a thorough overview of anti-Copernican astronomers and Brahe's ideas, see Kerry V. Magruder, "Jesuit Science After Galileo: The Cosmology of Gabriele Beati", *Centaurus* 51:3 (2009), 189-212.